**Title:** LARGE EDDY SIMULATION OF PREMIXED COMBUSTION WITH A

THICKENED-FLAMEAPPROACH

**Authors:**

Ashoke De[1], Graduate Student,
Sumanta Acharya [*,1,2], Professor,

[1]Mechanical Engineering Department
Louisiana State University, Baton Rouge, LA 70803

[2]Turbine Innovation and Energy Research Center
Louisiana State University, Baton Rouge, LA 70803

[*]Corresponding Author

Tel.: +1 225 578 5809 Fax: +1 225 578 5924

E-mail address: acharya@me.lsu.edu

**Abstract**

A Thickened Flame (TF) modeling approach is combined with a Large Eddy Simulation (LES) methodology to model premixed combustion and the accuracy of these model predictions is evaluated by comparing with the piloted premixed stoichiometric methane-air flame data of Chen et al. [Combust. Flame 107 (1996) 233-244] at a Reynolds number Re = 24,000. In the TF model, the flame front is artificially thickened to resolve it on the computational LES grid and the reaction rates are specified using reduced chemistry. The response of the thickened flame to turbulence is taken care of by incorporating an efficiency function in the governing equations. The efficiency function depends on the characteristics of the local turbulence and on the characteristics of the premixed flame such as laminar flame speed and thickness. Three variants of the TF model are examined: the original Thickened Flame model, the Power-law flame wrinkling model, and the dynamically modified TF model. Reasonable agreement is found when comparing predictions with the experimental data and with computations reported using a probability distribution function (PDF) modeling approach. The results of the TF model are in better agreement with data when compared with the predictions of the G-equation approach.

## Nomenclature

| | |
|---|---|
| A | pre-exponential constant |
| a,b | $\Gamma$ efficiency model exponents |
| $c_{ms}$ | material surface constant ~ 0.28 |
| $C_s$ | LES model coefficient |
| C | mean reaction progress variable |
| C' | rms of mean reaction progress variable |
| $D_i$ | molecular diffusivity |
| $D_{i,loc}$ | local molecular diffusivity |
| Da | Damköhler number |
| E | efficiency function |
| $E_a$ | activation energy |
| F | thickening factor |
| $F_{loc}$ | local thickening factor |
| $l_{F,t}$ | flame brush thickness |
| $l_t$ | integral length scale |
| $Re_t$ | turbulent Reynolds number |
| $Re_\Delta$ | sub-grid scale turbulent Reynolds number |
| $s_L^0$ | laminar flame speed |
| $S_{ij}$ | mean strain rate tensor |
| Sc | Schmidt number |
| $Sc_t$ | turbulent Schmidt number |
| $T_a$ | activation temperature |
| $T_b$ | adiabatic flame temperature |
| $T_u$ | temperature of unburnt mixtures |
| $u_i$ | velocity vector |
| $u'$ | rms turbulence velocity |
| $u'_{\Delta_e}$ | sub-grid scale turbulent velocity |
| $x_i$ | Cartesian coordinate vector |
| $Y_i$ | species mass fraction |

**Greek symbols**

| | |
|---|---|
| α | model constant |
| β | model constant |
| $\delta_L^0$ | laminar flame thickness |
| $\delta_L^1$ | thickened flame thickness |
| $\Delta_x$ | mesh spacing |
| $\Delta_e$ | local filter size |
| Γ | efficiency function |
| $\mu_t$ | dynamic turbulent eddy viscosity |
| $\nu_t$ | kinematic turbulent eddy viscosity |
| $\overline{\rho}$ | mean density |
| $\tau_c$ | chemical time scale |
| $\tau_i^k$ | sub-grid stress tensor |
| $\tau_t$ | turbulent time scale |
| $\omega_i$ | reaction rate |
| Ξ | wrinkling factor |

**Introduction**

The majority of the land based gas turbines are operated in a premixed mode due, in part, to environmental regulations for reducing NOx. To properly design premixed combustion systems, accurate predictions of premixed flames are desirable. Since the capability of the classical approach of using Reynolds averaged Navier-Stokes (RANS) equations in conjunction with phenomenological combustion models [1] is limited, numerical simulations of reacting flows based on large eddy simulations (LES) have been pursued since they are well suited to provide accurate and cost-effective predictions. The main philosophy behind LES of a reacting flow is to explicitly simulate the large scales of the flow and reactions, and to model the small scales. Hence, it is capable of capturing the unsteady phenomenon more accurately. The unresolved small scales or sub-grid scales must be modeled accurately to include the interaction between the turbulent scales. For premixed combustion, since the typical premixed flame thickness is smaller than the computational grid ($\Delta$), the small scale or sub-grid scale modeling must also take care of the interaction between turbulence and the combustion processes. In the Thickened Flame (TF) model [2] the flame is artificially thickened to resolve it on the computational mesh and to enable reaction rates from kinetic models to be specified in the flame region using reduced mechanisms. The influence of turbulence is represented by a parameterized efficiency function. A key advantage of the TF model is that it directly solves the species transport equations and uses the Arrhenius formulation for the evaluation of the reaction rates.

In the present paper, the TF model [2] is used, along with its variants, the Power-law flame wrinkling model [3, 4], and the dynamically modified version of these models [5]. A generic premixed combustion configuration is adopted for which experimental data [6] and numerical predictions with other approaches [7, 8] are available. These simulations and comparison with the benchmark data will help identify the accuracy of the TF modeling approach. In particular, earlier studies with this approach [2-4] have not provided a comparative assessment of the model and its variants with respect to detailed measurements, nor have the model predictions been compared with those from the more commonly used, but computationally more expensive, PDF approach, and the commonly used G-equation approach. This paper provides the first detailed comparative assessment (that includes velocity, turbulence kinetic energy, temperature, and major species) of all three modeling approaches and experimental data.

**Flow modeling using LES**

To model the turbulent flowfield, LES is used so that the energetic larger-scale motions are resolved, and only the small scale fluctuations are modeled. Therefore, the equations solved are the filtered governing equations for the conservation of mass, momentum, energy and species transport in curvilinear coordinate system [9-11]. The subgrid stress modeling is done using a dynamic Smagorinsky model where the unresolved stresses are related to the resolved velocity field through a gradient approximation:

$$\overline{u_i u_j} - \overline{u}_i \overline{u}_j = -2\nu_t \overline{S}_{ij} \tag{1}$$

$$\text{where} \quad \nu_t = C_s^2 (\Delta)^2 |\overline{S}| \tag{2}$$

$$\overline{S}_{ik} = \frac{1}{2}\left( (\vec{a}^m)_k \frac{\partial \overline{u}_i}{\partial \xi_m} + (\vec{a}^m)_k \frac{\partial \overline{u}_k}{\partial \xi_m} \right) \tag{3}$$

$$|\overline{S}| = \sqrt{2\overline{S}_{ik}\overline{S}_{ik}} \tag{4}$$

and S is the mean rate of strain. The coefficient $C_s$ is evaluated dynamically [12, 13] and locally-averaged.

**Combustion modeling**

Modeling the flame-turbulence interaction in premixed flames requires tracking of the thin flame front on the computational grid. Three major approaches have been used in the combustion modeling community.

(a) *G-equation model*: This approach treats the flame surface as an infinitely propagating surface (flamelet), and hence the flame thickness is treated approximately to zero [14, 15]. The propagating surface, which is called the flame front, is tracked using a field variable or iso-surface $G_o$. This technique is valid in both the corrugated flamelet regime and the thin reaction zone regime. However, the signed-distance function (which views the scalar G surrounding the front as the signed distance to the front) plays an important role since the G-equation only captures the instantaneous flame front. This dependence on the distance function is an inherent drawback of this method.

(b) *PDF approach*: The probability density function (pdf) is a stochastic method, which directly considers the probability distribution of the relevant stochastic quantities in a turbulent reacting flow. This kind of pdf description of turbulent reacting flow has some theoretical benefits, such as the complex chemistry is taken care of without applying any *ad-hoc* assumptions (like 'flamelet' or 'fast reaction'). Moreover, it can be applied to non-premixed, premixed, and partially premixed flames without having much difficulty. Usually, there are two ways which are mainly used to calculate the pdf: one is *presumed pdf approach*, and other is *pdf transport balance equation approach*. The *presumed pdf approach*, which essentially assumes the shape of the probability function P, is relatively simpler to use, however, has severe limitations in the context of applicability. On the other hand, the *pdf transport balance equation approach* solves a transport equation for pdf function, which is applicable for multi species, mass-weighted probability density function. This method has considerable advantage over any other turbulent combustion model due to its inherent capability of handling any complex reaction mechanism. However, the major drawback of transport pdf approach is its high dimensionality, which essentially makes the implementation of this approach to different numerical techniques, like FVM or FEM, limited, since their memory requirements increase almost exponentially with dimensionality. Usually, Monte-Carlo algorithms, which reduce the memory requirements, are used by Pope [16]. Moreover, a large number of particles need to be present in each grid cell to reduce the statistical error; however this makes it a very time consuming process. So far, the transport equation method has been only applied to relatively simple situations.

(c) *Thickened Flame model*: In this technique, the flame front is artificially thickened to resolve on computational grid. Corrections are made to ensure that the flame is propagating at the same speed as the unthickened flame [2, 3]. The key benefit of this approach, as noted earlier, rests in the ability to computationally resolve the reaction regions and the chemistry in these regions. More details on this approach are described in the following sections.

In the present paper we use the thickened flame approach due to its ability of using less computational resources and the model predictions are examined for a methane flame with detailed velocity and temperature data [6] available. As noted earlier, the paper will provide a detailed evaluation of the TF model relative to two other popular choices, the PDF model and the G-equation model.

**Thickened-Flame (TF) model:**

Butler and O'Rourke [17] were the first to propose the idea of capturing a propagating premixed flame on a coarser grid. The basic idea with this approach is that the flame is artificially thickened to include several computational cells and by adjusting the diffusivity to maintain the same laminar flame speed $s_L^0$. It is well known from the simple theories of laminar premixed flames [18, 19] that the flame speed and flame thickness can be related through the following relationship

$$s_L^0 \propto \sqrt{D\overline{B}}, \delta_L^0 \propto \frac{D}{s_L^0} = \sqrt{\frac{D}{\overline{B}}} \tag{5}$$

where D is the molecular diffusivity and $\overline{B}$ is the mean reaction rate. When the flame thickness is increased by a factor F, the molecular diffusivity and reaction rate are modified accordingly

(FD and $\bar{B}/F$) to maintain the same flame speed. The major advantages associated with thickened flame modeling are: (i) the thickened flame front which is resolved on LES mesh, is usually less than typical premixed flame thickness (around 0.1-1 mm), (ii) quenching and ignition events can be simulated, (iii) chemical reaction rates are calculated exactly like in a DNS calculation without any *ad-hoc* sub models so that it can theoretically be extended to incorporate multi-step chemistry [2].

In LES framework, the spatially filtered species transport equation can be written as

$$\frac{\partial \overline{\rho} Y_i}{\partial t} + \frac{\partial}{\partial x_j}(\overline{\rho} Y_i u_j) = \frac{\partial}{\partial x_j}\left(\overline{\rho D_i \frac{\partial Y_i}{\partial x_j}}\right) - \frac{\partial}{\partial x_j}\left[\overline{\rho}\left(\widetilde{Y_i u_j} - Y_i u_j\right)\right] + \overline{\dot{\omega}_i} \tag{6}$$

where the terms on the r.h.s. are the filtered diffusion flux, the unresolved transport, and the filtered reaction rate respectively. In general, the unresolved term is modeled with a gradient diffusion assumption by which the laminar diffusivity is augmented by turbulent eddy diffusivity. However, in the TF model, the "thickening" procedure multiplies the diffusivity term by a factor F which has the effect of augmenting the diffusivity. Therefore, the gradient approximation for the unresolved fluxes is not explicitly used in the closed species transport equations. The corresponding filtered species transport equation in the thickened-flame model becomes

$$\frac{\partial \overline{\rho} Y_i}{\partial t} + \frac{\partial}{\partial x_j}(\overline{\rho} Y_i u_j) = \frac{\partial}{\partial x_j}\left(\overline{\rho} F D_i \frac{\partial Y_i}{\partial x_j}\right) + \frac{\overline{\dot{\omega}_i}}{F} \tag{7}$$

Although the filtered thickened flame approach looks promising, a number of key issues need to be addressed. The thickening of the flame by a factor of F modifies the interaction between turbulence and chemistry, like the Damköhler number, Da, which is a ratio of the turbulent ($\tau_t$)

and chemical ($\tau_c$) time scales, is decreased by a factor F and becomes Da/F, where

$$Da = \frac{\tau_t}{\tau_c} = \frac{l_t s_L^0}{u' \delta_L^0} \tag{8}$$

As the *Da* is decreased, the thickened flame becomes less sensitive to turbulent motions. Therefore, the sub-grid scale effects have been incorporated into the thickened flame model, and parameterized using an efficiency function E derived from DNS results [2, 3]. Using the efficiency function, the final form of species transport equation becomes

$$\frac{\partial \overline{\rho} Y_i}{\partial t} + \frac{\partial}{\partial x_j}(\overline{\rho} Y_i u_j) = \frac{\partial}{\partial x_j}\left(\overline{\rho} EFD_i \frac{\partial Y_i}{\partial x_j}\right) + \frac{E\overline{\dot{\omega}_i}}{F} \tag{9}$$

where the modified diffusivity ED, before multiplication by F to thicken the flame front, may be decomposed as ED=D(E-1)+D and corresponds to the sum of molecular diffusivity, D, and a turbulent sub-grid scale diffusivity, (E-1)D. In fact, (E-1) D can be regarded as a turbulent diffusivity used to close the unresolved scalar transport term in the filtered equation.

*The original TF model*

The central ingredient of the TF model is the sub-grid scale wrinkling function E, which is defined by introducing a dimensionless wrinkling factor Ξ. The factor Ξ is the ratio of flame surface to its projection in the direction of propagation. The efficiency function, E, is written as a function of the local filter size ($\Delta_e$), local sub-grid scale turbulent velocity ($u'_{\Delta_e}$), laminar flame speed ($s_L^0$), and the thickness of the laminar and the artificially thickened flame ($\delta_L^0, \delta_L^1$). Colin et al. [2] proposed the following expressions for modeling the efficiency function

$$\Xi = 1 + \beta \frac{u'_{\Delta_e}}{s_L^0} \Gamma\left(\frac{\Delta_e}{\delta_L^0}, \frac{u'_{\Delta_e}}{s_L^0}\right) \tag{10}$$

$$\beta = \frac{2\ln 2}{3c_{ms}\left(\text{Re}_t^{1/2}-1\right)}, c_{ms} = 0.28, \text{Re}_t = u'l_t/\nu \tag{11}$$

where $\text{Re}_t$ is the turbulent Reynolds number. The local filter size $\Delta_e$ is related with laminar flame thickness as

$$\Delta_e = \delta_L^1 = F\delta_L^0 \tag{12}$$

The function $\Gamma$ represents the integration of the effective strain rate induced by all scales affected due to artificial thickening, $\Gamma$ is estimated as

$$\Gamma\left(\frac{\Delta_e}{\delta_L^0}, \frac{u'_{\Delta_e}}{s_L^0}\right) = 0.75 \exp\left[-1.2\left(\frac{u'_{\Delta_e}}{s_L^0}\right)^{-0.3}\right]\left(\frac{\Delta_e}{\delta_L^0}\right)^{2/3} \tag{13}$$

Finally, the efficiency function takes the following form as defined by the ratio between the wrinkling factor, $\Xi$, of laminar flame ($\delta_L = \delta_L^0$) to thickened flame ($\delta_L = \delta_L^1$).

$$E = \frac{\Xi|_{\delta_L = \delta_L^0}}{\Xi|_{\delta_L = \delta_L^1}} \geq 1 \tag{14}$$

where the subgrid scale turbulent velocity is evaluated as $u'_{\Delta_e} = 2\Delta_x^3 \left|\nabla^2\left(\nabla \times \bar{u}\right)\right|$, and $\Delta_x$ is the grid size. This formulation for sub-grid scale velocity estimation is free from dilatation. Usually, $\Delta_e$ differs from $\Delta_x$, and it has been suggested that values for $\Delta_e$ be at least $10\Delta_x$ [2].

*Power-law flame wrinkling model*

This model uses a different estimation of efficiency function that was proposed by

Charlette et al. [3, 4]. This approach basically relates the flame surface area to a cutoff length which limits wrinkling at the smallest length-scales of the flame. Based on asymptotic analysis, the efficiency function is evaluated using the following relationships.

$$E = \left(1 + \min\left[\frac{\Delta_e}{\delta_L^0}, \Gamma \frac{u'_{\Delta_e}}{s_L^0}\right]\right)^\alpha \tag{15}$$

where $\Gamma$ is defined as:

$$\Gamma\left(\frac{\Delta_e}{\delta_L^0}, \frac{u'_{\Delta_e}}{s_L^0}, \text{Re}_\Delta\right) = \left[\left(\left(f_u^{-a} + f_\Delta^{-a}\right)^{-1/a}\right)^{-b} + f_{\text{Re}}^{-b}\right]^{-1/b} \tag{16}$$

$$f_u = 4\left(\frac{27 C_k}{110}\right)\left(\frac{18 C_k}{55}\right)\left(\frac{u'_{\Delta_e}}{s_L^0}\right)^2 \tag{17}$$

$$f_\Delta = \left[\frac{27 C_k \pi^{4/3}}{110} \times \left(\left(\frac{\Delta_e}{\delta_L^0}\right)^{4/3} - 1\right)\right]^{1/2} \tag{18}$$

$$f_{\text{Re}} = \left[\frac{9}{55}\exp\left(-\frac{3}{2} C_k \pi^{4/3} \text{Re}_\Delta^{-1}\right)\right]^{1/2} \times \text{Re}_\Delta^{1/2} \tag{19}$$

The constants a, b and $C_k$ control the sharpness of the transitions between the asymptotic behaviors. The suggested values are b=1.4, $C_k$=1.5,

$$a = 0.6 + 0.2\exp\left[-0.1\left(u'_{\Delta_e}/s_L^0\right)\right] - 0.20\exp\left[-0.01\left(\Delta_e/\delta_L^0\right)\right]$$
$$\Delta_e = F\delta_L^0, \text{Re}_\Delta = 4\frac{\Delta_e}{\delta_L^0}\frac{u'_{\Delta_e}}{s_L^0} \tag{20}$$

In present work, $\alpha$=0.5 is used, resulting the non-dynamic formulation for Power-law model.

*Dynamically modified TF model*

The major issue associated with TF model is that it modifies the diffusion term

(multiplying with a thickening factor F) throughout the whole computational domain. This may lead to inaccuracies in the species mass fraction prediction, especially in the case of inhomogeneous premixed or partially premixed combustion. To overcome this disadvantage, a dynamically thickened flame model has initially been developed at CERFACS [20, 21]. Recently, Durand and Polifke [5] came up with a different expression for the diffusivity that is based on the reaction progress variable, and is used in the present work in the following form:

$$F_{loc} = 1 + (F-1)\Omega(c) \tag{21}$$

$$\Omega(c) = 16\left[c(1-c)\right]^2, c = 1 - \frac{Y_F}{Y_F^{st}} \tag{22}$$

$$D_{i,loc} = \frac{\mu}{Sc} EF_{loc} + (1-F_{loc})\frac{\mu_t}{Sc_t} \tag{23}$$

As seen in the above equation, the global thickening factor F and diffusivity D are both represented in terms of a local function $\Omega(c)$, evaluated based on reaction progress variable. Hence, the species transport equation is also modified and takes the form:

$$\frac{\partial \overline{\rho Y_i}}{\partial t} + \frac{\partial}{\partial x_j}(\overline{\rho Y_i u_j}) = \frac{\partial}{\partial x_j}\left(\overline{\rho}EF_{loc}D_{i,loc}\frac{\partial Y_i}{\partial x_j}\right) + \frac{E\overline{\dot{\omega}_i}}{F_{loc}} \tag{24}$$

All of these afore-mentioned models are used in the present work in conjunction with 1 and 2 step global Arrhenius chemistry.

**Chemistry model**

As all the species are explicitly resolved on the computational grid, the Thickened Flame model is best suited to resolve major species. Intermediate radicals with very short time scales can not be resolved. To this end, only simple global chemistry has been used with the thickened flame model. In the present study, 1 and 2 step reaction chemistry are explored.

A single step scheme, which includes five species ($CH_4$, $O_2$, $H_2O$, $CO_2$ and $N_2$) is given by the following expression.

$$CH_4 + 2O_2 \rightarrow CO_2 + 2H_2O$$

where the reaction rate expression is given by

$$q_1 = A\exp(-T_a/T)[CH_4]^a[O_2]^b \tag{25}$$

The activation temperature $T_a$ is 24,358 K, parameters a=0.2, b=1.3, and pre-exponential factor $A=2.29 \times 10^{13}$ as given by Kim et al. [22].

A two step chemistry, which includes six species ($CH_4$, $O_2$, $H_2O$, $CO_2$, CO and $N_2$) is given by the following equation set.

$$CH_4 + 1.5O_2 \rightarrow CO + 2H_2O$$

$$CO + 0.5O_2 \leftrightarrow CO_2$$

The corresponding reaction rate expressions are given by:

$$q_1 = A_1 \exp(-E^1_a/RT)[CH_4]^{a1}[O_2]^{b1} \tag{26}$$

$$q_2(f) = A_2 \exp(-E^2_a/RT)[CO][O_2]^{b2} \tag{27}$$

$$q_2(b) = A_2 \exp(-E^2_a/RT)[CO_2] \tag{28}$$

where the activation energy $E^1_a = 34500$ cal/mol, $E^2_a = 12000$ cal/mol, a1=0.9, b1=1.1, b2=0.5, and $A_1$ and $A_2$ are 2.e+15 and 1.e+9, respectively, as given by Selle et al. [23]. Properties including density of mixtures are calculated using CHEMKIN-II [24] and TRANFIT [25] depending on the local temperature and the composition of the mixtures at 1 atm.

**Solution Procedure**

In the present study, a parallel multi-block compressible flow code for an arbitrary number of reacting species, in generalized curvilinear coordinates is used. Chemical mechanisms and thermodynamic property information of individual species are input in standard Chemkin format. Species equations along with momentum and energy equation are solved implicitly in a fully coupled fashion using a low Mach number preconditioning technique, which is used to effectively rescale the acoustics scale to match that of convective scales [26]. An Euler differencing for the pseudo time derivative and second order backward 3-point differencing for physical time derivatives are used. A second order low diffusion flux-splitting algorithm is used for the convective terms [27]. However, the viscous terms are discretized using second order central differences. An incomplete Lower-Upper (ILU) matrix decomposition solver is used. Domain decomposition and load balancing are accomplished using a family of programs for partitioning unstructured graphs and hypergraphs and computing fill-reducing orderings of sparse matrices, METIS. The message communication in distributed computing environments is achieved using Message Passing Interface, MPI. The multi-block structured curvilinear grids presented in this paper are generated using commercial grid generation software GridPro$^{TM}$.

**Flow configuration**

The configuration of interest in the present work is the Bunsen burner geometry investigated by Chen et al. [6], is shown in Figure 1. The flame is a stoichiometric premixed methane-air flame, stabilized by an outer pilot. The incoming streams of both the main and pilot jets have the same composition. The main jet nozzle diameter (D) is 12 mm. The pilot stream is supplied through a perforated plate (1175 holes of 1 mm in diameter) around the main jet, with an outer diameter of 5.67D. The Reynolds number used in the present work is Re=24,000 (flame F3 in [6]). Based on the estimated characteristic length and time scale given in Chen et al. [6], $u'/S_L$=11.9, $l/l_f$=13.7, flame F3 corresponds to the thin reaction zone regime.

The computational domain extends 20D downstream of the fuel-air nozzle exit, 4D upstream of the nozzle exit and 4D in the radial direction. Two different LES grids are studied

(for cold flow only): one that consists of 300x94x64 grid points downstream of the nozzle exit plus 50x21x64 grid points upstream, and corresponds to approximately 1.88M grid points (mesh1: coarse). The finer mesh consists of 444x140x96 grid points downstream of the nozzle plus 74x31x96 grid points upstream, and contains approximately 5.91M grid points (mesh2: fine).

At the inflow boundary, the instantaneous velocities are computed using a random flow generation technique [28]. Convective boundary conditions [29] are prescribed at the outflow boundary, and stress-free conditions are applied on the lateral boundary in order to allow the entrainment of fluid into domain. The time step used for the computation is dt=1.0e-3, and the heated pilot temperature is chosen 2005K. The inlet temperature of the main premixed fuel-air jet and the initial temperature in the calculation domain is specified as 300K.

**Non-reacting Flow Results**

In order to validate the flow solver, LES calculation for the non-reacting flow is reported in this section. That ensures that the grid and boundary conditions are properly chosen, and that the subgrid model is able to capture the modeled turbulence adequately.

Figure 2(a) represents the instantaneous snap shot of the velocity fields. The evolution of three components of velocities is clearly observed and shows that the radial and tangential velocities are of the same order of magnitude, and considerably lower in magnitude relative to the peak axial velocity. The premixed-jet breakup appears to occur about 3-4 jet-diameters downstream of injection.

*Mean axial velocity*

The mean axial velocity profiles in the radial directions at different axial locations are shown in Fig. 3. The time-averaged mean velocity is normalized by the bulk velocity $U_o$=30 m/s.

As observed from the experimental data and predictions, the jet shows the expected spreading behavior. The potential core appears to persist up to about X/D=4.5, beyond which the centerline mean velocity decreases as the jet expands in the radial direction. These effects are well reproduced by the LES simulations, and the overall predictions are in good agreement with the experimental results. Moreover, the coarse mesh (1.88M grid points) results are in good agreement with the fine mesh (5.91M grid points), and indicate that the 1.88M node calculations are grid-independent.

*Turbulent kinetic energy*

The radial profile of the turbulent kinetic energy, normalized by square of the bulk velocity, at different axial locations is shown in Fig. 3. It is evident from the figure that the computed results are reasonably well predicted and compare well with the experimental data set. At X/D=4.5, the center line peak value for the fine mesh is slightly over predicted, but at X/D=6.5 this over-prediction is eliminated. The evolution of kinetic energy follows the expected trend and initially shows the kinetic energy peak in the mixing layer formed by the primary-jet and the coaxial air-stream. Further downstream, the mixing layers merge and the kinetic energy peak moves toward the centerline as seen at X/D=8.5.

**Reacting Flow Results-Assessment of Model Variants**

Figure 2(b) shows a typical evolution of instantaneous flame front, which is represented by the temperature field. As time increases the flame front is seen to expand radially and in the downstream directions with the burnt regions having higher flame temperatures, and the unburnt regions (in blue) at lower temperatures. The cusp formation (regions with negative curvature) is clearly observed from the instantaneous field along the boundaries of the burnt (product side) and unburnt (air or premixed fuel-air side) regions. The cusps are usually formed towards the product side, where the flow field is accelerated due to the heat release from the flame. However, the small scale cusp formation is reduced by thickening the flame front artificially (seen as narrow band in Fig 2(b)), and this reduces the flame wrinkling as well.

*Comparison of the TF model and its Variants*

Figure 4 shows the predicted results for variants of TF model using 2-step and 1-step chemistry. It is observed that the various TF model predictions do not show any major differences. The dynamic versions of the models do show some improvements, while the power-law model for the two-step chemistry model also shows an improvement in the prediction of the peak kinetic energy. However, the profiles with all the models are relatively close to each other and no one model consistently shows improved prediction over the others at the various X/D locations. Hence, based on this comparison, the original TF model would be an appropriate choice, and therefore, further calculations reported are based on the original TF model. It is also observed that the velocity predictions using the G-eqn. model do not show good agreement with the measurements while both the TF model and the PDF model predictions are in good agreement with the mean velocity data. The turbulent kinetic energy predictions with the PDF model appear to be over-predicted at X/D=2.5 at all radial locations, while the TF model tends to under-predict at radial locations close to centerline.

*Chemistry model comparison*

The mean temperature profiles obtained from different chemistry models are presented in Figure 5. The mean progress variable is defined as $C= (T-T_u)/(T_b-T_u)$, where $T_u$= 298K and $T_b$=2248 K.

It is observed that the 2-step chemistry predictions appear to be a significant improvement over 1-step chemistry predictions and agree better with experimental data.

Moreover, the different variants of the TF model with 2-step chemistry produce very comparable predictions. Hence, from this point we will present original TF model predictions using 2-step chemistry. It should be noted that the G-equation model predictions generally over-predict the temperature values significantly and appear to perform much worse than the PDF or TF models.

**Original TF model Results-Detailed Distributions**

Based on the observations so far that show improvements with the two-step chemistry calculations, and no clear difference between the variants of the TF models, additional results will be presented only with the original TF model and with the two-step chemistry calculations. In this section, attention will be focused on detailed comparisons of the TF model, the PDF model, the G-equation model and the experimental data for the mean velocity, temperature, kinetic energy and all major species. This comparison will be presented at several X/D locations (2.5, 4.5, 6.5, and 8.5).

*Mean axial velocity*

Figure 6 shows the mean axial velocity profiles at different axial locations obtained using 2-step global chemistry. Also shown are the predictions of Lindstedt et al. [7] (used PDF model for his calculations) and Duchamp et al. [8] (used G-equation for his simulations).

In evaluating the models, the perspective of computational economy must be kept in mind. A multi step calculation require sthe calculation of transport equations for multiple species, and with a PDF modeling approach, the reaction rate expressions require the use of a look-up table with substantial computational input/output (I/O) requirements. In the Thickened Flame modeling approach, the reaction rate expressions are computed using Arrhenius law, and computational effort including I/O time is expected to be less. However, we have not undertaken a direct study of the two approaches comparing the computational efforts at this stage.

Figure 6 shows that the mean velocity predictions using the TF modeling approach and these are in reasonably good agreement with experimental data as well as with the PDF model predictions. At the downstream locations, the TF model predictions show higher spreading along the inner edge of the mixing layer (the PDF model shows a similar behavior also). In comparing the models, the G-eqn shows significant over-prediction with the data for X/D<6.5.

The radial profiles of the mean axial velocity show greater radial broadening compared to the non-reacting case (as shown in Fig. 3), due the effect of the flame front, pushing the shear layer outward in the radial direction. Furthermore, it is observed that the peak center line velocity remains almost constant in the axial direction, and exhibits a longer potential core compared to the non-reacting case. These effects are reasonably well reproduced by the simulations.

*Turbulent kinetic energy*

Turbulent kinetic energy predictions using the TF modeling approach are shown in Figure 7. As it is evident from the experimental data there are significant differences observed while compared to the non-reacting case (Fig. 3). The measured turbulent kinetic energy initially increases with axial distance for the reacting case while it decreases for the non-reacting case. Furthermore, for the reacting case, the kinetic energy peak moves away from the centerline (due to radial expansion of the flame front) with increasing axial direction while the peak moves towards centerline in the non-reacting case due to turbulent diffusion effects.

The predicted kinetic energy shows significant differences between the various models. The TF model shows an initial under-prediction close to the centerline (at X/D=2.5), but the peak values and the general trend is well predicted by the TF model. At the downstream locations (X/D=4.5 and 6.5), the TF model predictions are in reasonably good agreement with the experimental data. The PDF model over predicts the turbulence levels at all radial locations in the near field of injection, but appears to be in better agreement with the measured data at X/D=8.5. The G-equation model generally shows the poorest performance excluding the near-injection location at X/D=2.5 where the agreement is reasonable. These discrepancies are linked with the mean temperature predictions and discussed in the following sub-section. Clearly, in assessing all X/D locations, the TF model predictions appear to provide the best agreement with the data.

*Mean temperature*

Mean temperature profiles are presented in Figure 8. Immediately downstream of the nozzle exit (X/D=2.5), the temperature is over-predicted due to the effect of the 2-step chemistry model (as opposed to more complex chemistry schemes) and is likely to over-predict temperature in fuel-rich regions. However, further downstream (X/D=4.5, X/D=6.5) the peak temperature predictions with the TF model and the PDF models are in better agreement with the data. In particular, the TF model and the PDF model show good agreement, while the G-equation consistently shows an over prediction. At X/D=8.5 all models excluding the G-equation under-predict the measured temperature distribution. With the TF model, it is apparent that the spreading of the temperature shear layer is not being correctly predicted, and that this is likely due to the artificial thickening of the flame front and the scalar diffusivity in the flame region. It can be seen that the centerline mean temperature increases downstream in the experiments, whereas it remains close to the unburnt temperature in the TF simulation. While both the PDF model and the TF model show qualitatively similar behavior, the PDF model does show somewhat better agreement with the temperature data. This is a consequence of the more detailed chemistry calculations incorporated in their models. Despite this, the poorer predictions of the turbulent kinetic energy with the PDF model, relative to the TF model, are puzzling.

*Species mass fraction*

The radial distributions of the major species mass fractions are presented in the Figures 9-10. In presenting these results, the G-equation model calculations are not presented here since the detailed data with this model is not available.

Figure 9 shows the mean $CH_4$ distributions at the four radial locations. The profile of the curve clearly shows the premixed unburnt core (plateau region near the centerline), the flame region where methane is consumed (represented by the decay region), and the burnt or outer region where no methane is present. It can be seen that only by X/D=8.5 does the flame region start approaching the centerline. The TF model predictions are in excellent agreement with the data, and the level of agreement is even better than that observed with the PDF model. The $O_2$ concentrations are shown in Fig. 9, exhibit distributions similar to the $CH_4$ curve, and the experimental trends are well captured by both the TF model the PDF model. The TF model appears to over-predict oxygen concentrations in the outer regions. This observation leads to an under-prediction of $H_2O$ in the outer regions as well (Fig. 9).

$CO_2$ concentrations are also in good agreement with the measurements (Fig. 10) at all X/D locations, while CO predictions (Fig. 10) show excellent agreement at X/D of 2.5 but under-predict at other X/D locations. This may possibly be due to the lower $O_2$ consumption. However, PDF model simulations always significantly over estimate the CO concentrations, and also underestimate $CO_2$ concentrations. At this stage, the discrepancies in CO predictions are not very clear; hence one needs to be more careful in interpreting the data.

*Turbulent flame brush thickness*

The turbulent flame-brush thickness, which is a characteristic representation of the transition zone between burnt and unburnt gases in premixed flames, is computed as

$$l_{F,t} = \left[\frac{\partial C}{\partial r}\right]_{max}^{-1} \qquad (29)$$

Figure 11 represents the comparison of turbulent flame-brush thickness for TF model, with respect to experiments and other simulations. Over-prediction of flame-brush thickness inherently explains the under-prediction of mean reaction progress variable $C$, which is evident from Figure 8. Among all the models, TF model predictions and PDF simulations are in better agreement with the data.

**Conclusions**

In the present paper, a Thickened Flame approach is used to compute the piloted premixed stoichiometric methane-air flame for Reynolds number Re = 24,000. The original TF model and its variants including the Power-law flame wrinkling model, and the dynamically modified version of these models in conjunction with two different chemistry models have been implemented and compared. In the first part of the work, the comparison between various TF models and different chemistry models has been presented.

It has been shown that the various variants of TF model produce close agreement with each other using 2-step chemistry model. However, they do exhibit some differences while using 1-step chemistry. Moreover, the 2-step chemistry model appears to perform better over 1-step chemistry model. In the second part of the paper the detailed results using the original TF model with 2-step chemistry have been reported.

The original TF model predictions with 2-step chemistry have been found to be in satisfactory agreement with the experimental data and with the more detailed PDF simulations. The mean reaction progress variable and the mean axial velocity are well predicted in the near-field while showing some discrepancies at the downstream locations. The turbulent kinetic energy is under-predicted in the vicinity of the centerline at axial locations X/D=2.5, but matches well at the locations X/D=4.5-8.5. The major species mass fraction predictions are also in good agreement with the exception of CO that is under-predicted by the TF model and over-predicted by the PDF model. In general, the TF model and the previously published PDF model predictions are in reasonable agreement with each other and experiments, whereas the G-equation model predictions show poor performance.

In comparing the computational resources, the TF model is always been less computationally expensive compared to PDF model where the use of a look-up table takes substantial computational input/output (I/O) requirements. However, TF model simulations at this stage are restricted to very few number of chemical kinetics (1 or 2-step chemistry) while producing reasonably good agreement of the data.

**Acknowledgements**

This work was supported by the Clean Power and Energy Research Consortium (CPERC) of Louisiana through a grant from the Louisiana Board of Regents. Partial support for this work is also provided through Industrial Ties Research Subprogram from the Board of Regents. Simulations are carried out on the computers provided by LONI network at Louisiana, USA (www.loni.org) and HPC resources at LSU, USA (www.hpc.lsu.edu). This support is gratefully acknowledged.

**References**

1. Poinsot, T., Veynante, D., (2001). Theoretical and Numerical Combustion, Edwards.
2. Colin, O., Ducros, F., Veynante, D., Poinsot, T., (2000). A thickened flame model for large eddy simulation of turbulent premixed combustion. Physics of Fluids, 12(7), pp. 1843-1863.

3. Charlette, F., Meneveau, C., Veynante, D., (2002). A Power-Law flame wrinkling model for LES of premixed turbulent combustion, Part I: Non Dynamic formulation and initial tests. Combustion and Flame, 131, pp. 159-180.

4. Charlette, F., Meneveau, C., Veynante, D., (2002). A Power-Law flame wrinkling model for LES of premixed turbulent combustion, Part II: Dynamic formulation and initial tests. Combustion and Flame, 131, pp. 181-197.

5. Durand, L., Polifke, W., (2007). Implementation of the thickened flame model for large eddy simulation of turbulent premixed combustion in a commercial solver. ASME Turbo Expo 2007: Land, Sea and Air, ASME paper GT2007-28188.

6. Chen, Y. C., Peters, N., Schneemann, G. A., Wruck, N., Renz, U., Mansour, M. S., (1996). The detailed flame structure of highly stretched turbulent premixed methane-air flames, Combustion and Flame, 107, pp. 233-244.

7. Lindstedt, R. P., Vaos, E. M., (2006). Transported PDF modeling of high-Reynolds-number premixed turbulent flames, Combustion and Flame, 145, pp. 495-511.

8. Duchamp de La Geneste, L., Pitsch, H., (2000). A level-set approach to large-eddy simulation of premixed turbulent combustion, Annual Research Briefs, CTR, Stanford, pp. 105-116.

9. Jordan, Stephen A., (1999). A large-eddy simulation methodology in generalized curvilinear coordinates, Journal of Computational Physics, 148, pp. 322-340.

10. Jordan, Stephen A., (2001). Dynamic subgrid-scale modeling for large-eddy simulation in complex tropologies, Journal of Fluids Engineering, 123, pp. 619-627.

11. Tafti, D. K., (2005). Evaluating the role of subgrid stress modeling in a ribbed duct for the internal cooling of turbine blades, Int. J. Heat and Fluid Flow, 26, pp. 92-104.

12. Smagorinsky, J., (1963). General circulation experiments with the primitive equations. I: The basic experiment, Monthly Weather Review, 91, pp. 99-165.

13. Germano, M., Piomelli, U., Moin, P., Cabot, W. H., (1991). A dynamic subgrid-scale eddy viscosity model, Physics of Fluids, 3, pp. 1760-1765.

14. Peters, N., (2000). Tubulent Combustion, Cambridge Univ. Press, London/New York.

15. Pitsch, H., Duchamp de La Geneste, L., (2002). Large-eddy simulation of a premixed turbulent combustion using level-set approach, Proceedings of the Combustion Institute, 29, pp. 2001-2008.

16. Pope, S. B., (1985). Pdf methods for turbulent reactive flows, Progress in Energy Combustion Science, 11, pp. 119-192.

17. Butler, T. D., O'Rourke, P. J., (1977). A numerical method for two-dimensional unsteady reacting flows, Proceedings of the Combustion Institute, 16, pp. 1503-1515.

18. Kuo, Kenneth K., (2005). Principles of Combustion, Second edition. John Wiley & Sons. Inc.

19. Williams, F. A., (1985). Combustion Theory, Benjamin/Cummins, Menlo park, CA.

20. Legier, J. P., Poinsot, T., Veynante, D., (2000). Dynamically thickened flame LES model for premixed and non-premixed turbulent combustion, Summer Program, Center for Turbulent Research, Stanford University, pp. 157--168.

21. Tuffin, K., Varoquié, B., Poinsot, T., (2003). Measurements of transfer functions in reacting flows using Large eddy simulations, In 10th International Congress On Sound And Vibration, pp. 785--793.

22. Kim, N.I., Maruta, K., (2006). A numerical study on propagation of premixed flames in small tubes, Combustion and Flame, 146, pp. 283-301.

23. Selle, L., Lartigue, G., Poinsot, T., Koch, R., Schildmacher, K. U., Krebs, W., Prade, B., Kaufmann, P., Veynante, D., (2004). Compressible large eddy simulation of turbulent combustion in complex geometry on unstructured meshes, Combustion and Flame, 137, pp. 489-505.

24. Kee., R. J., Rupley, F. M., Miller, J. A., (1989). Chemkin-II: A Fortran Chemical Kinetics Package for Analysis of Gas-Phase Chemical Kinetics, Sandia Report, 89-8009B.

25. Kee, R. J., Dixon-Lewis, Warnats, J., Coltrin, M. E., Miller, J. A., (1986). Technical Report SAND86-8246 (TRANFIT), Sandia National Laboratories, Livermore, CA.

26. Weiss, J. M., Smith, W. A., (1995). Preconditioning applied to variable and constant density flows. AIAA Journal, 33, pp. 2050-2057.

27. Edwards, J. R., (1997). A low-diffusion flux-splitting scheme for Navier-Stokes calculations, Computers and Fluids, 26, pp. 635-659.

28. Smirnov, A., Shi, S., Celik, I., (2001). Random Flow generation technique for large eddy simulations and particle-dynamics modeling, J. of Fluids Engineering, 123, pp. 359-371.

29. Akselvoll, K., Moin, P., (1996). Large-eddy simulation of turbulent confined coannular jets, J. of Fluid Mechanics, 315, pp. 387-411.

30. Herrmann, M., (2006). Numerical simulation of turbulent Bunsen flames with a level set flamelet model, Combustion and Flame,145, pp. 357-375.

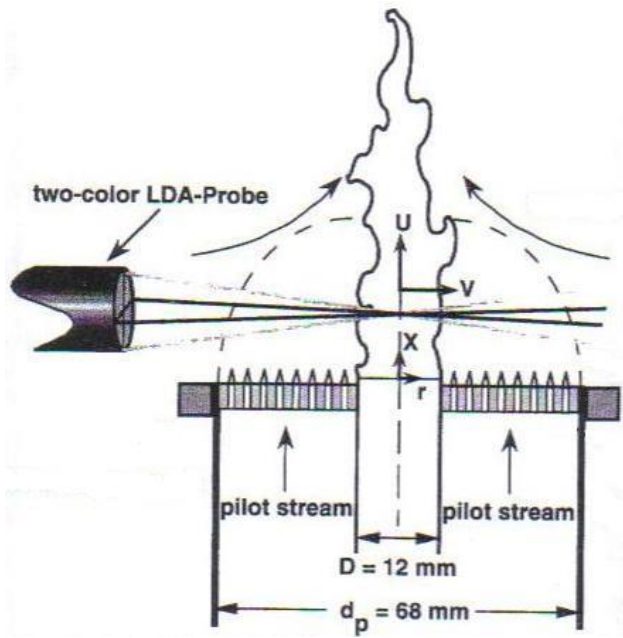

Figure 1. Schematic diagram of the Bunsen burner with enlarge pilot flame [6].

(a)

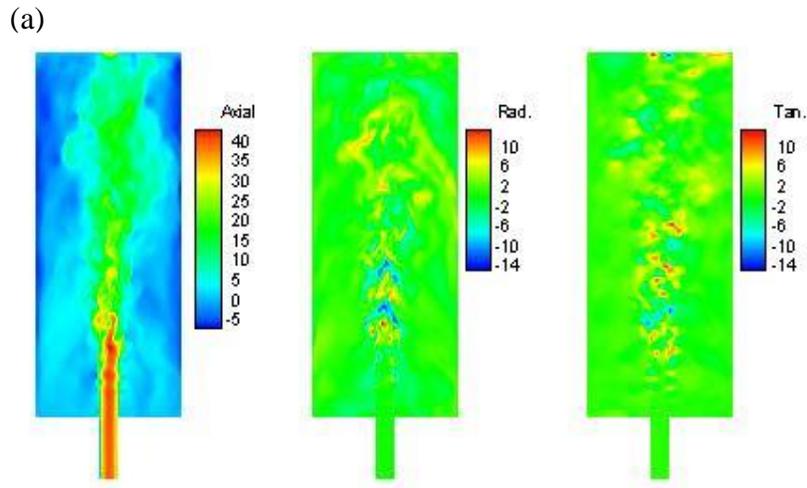

(b)

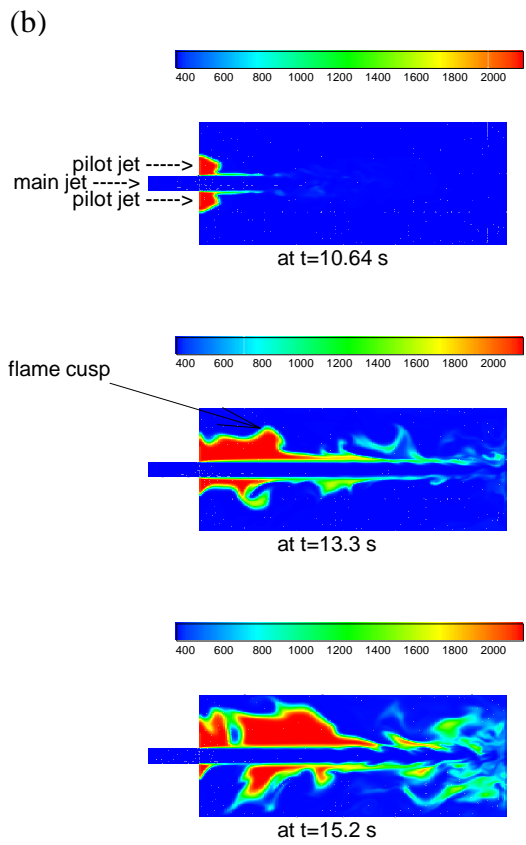

Figure 2. Instantaneous flow field. (a) Velocity (m/s), (b) Temperature ( K).

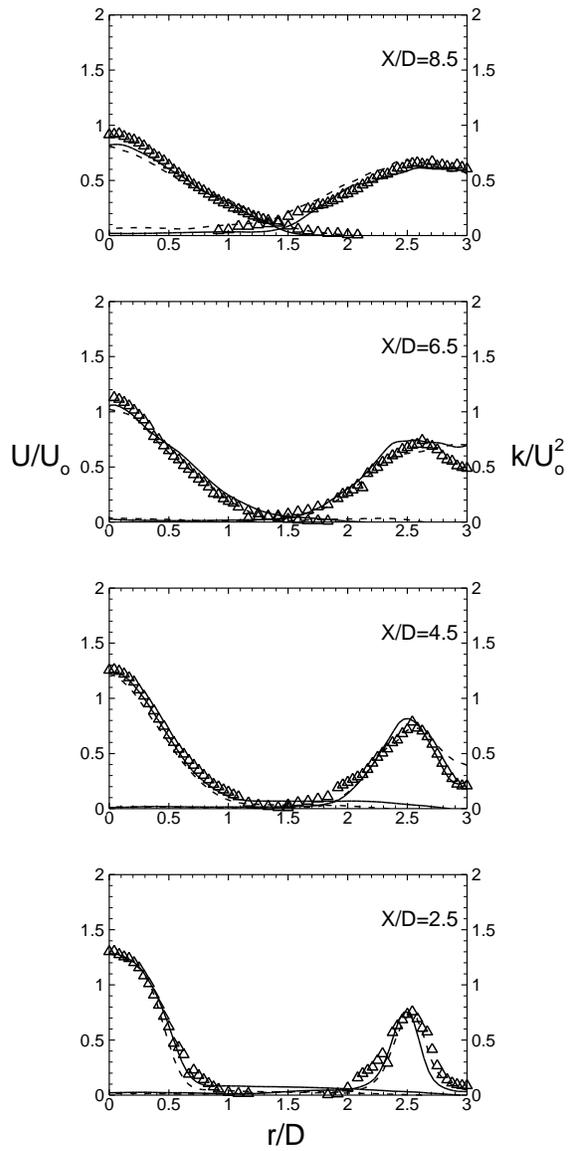

Figure 3. Cold flow: mean axial velocity $U/U_o$ and turbulent kinetic energy $k/U^2_o$X20. Experimental data is shown by symbols ($\Delta$) and lines are LES results: coarse (———), fine (- - -).

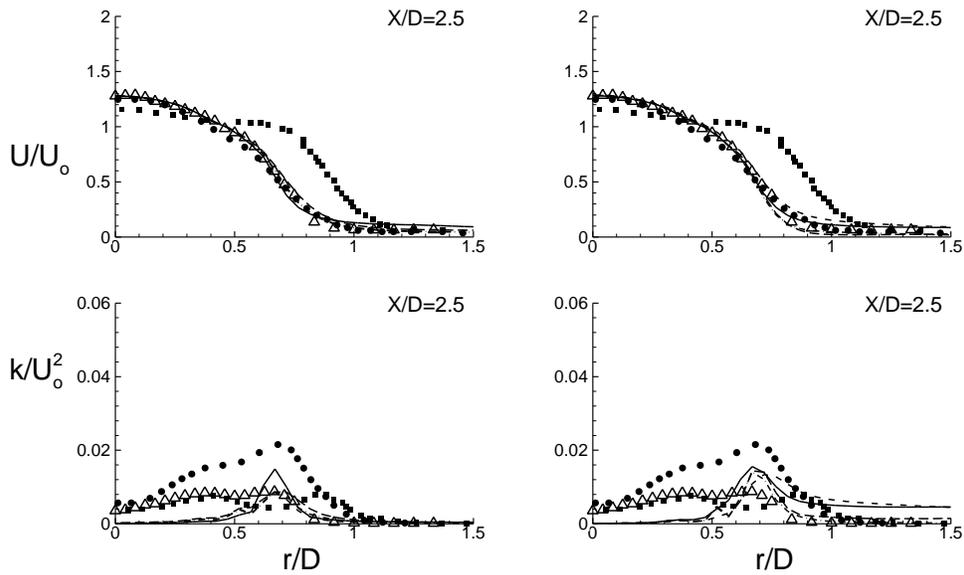

Figure 4. 2 step chemistry (left), 1 step chemistry (right): mean axial velocity $U/U_o$ (top), turbulent kinetic energy $k/U^2_o$ (bottom). Experimental data (Δ), Lindstedt simulations (●), Duchamp simulations (■), TF model (——), Power-law (----), Dynamically modified TF model (·····), Dynamically modified Power-law model (- - -).

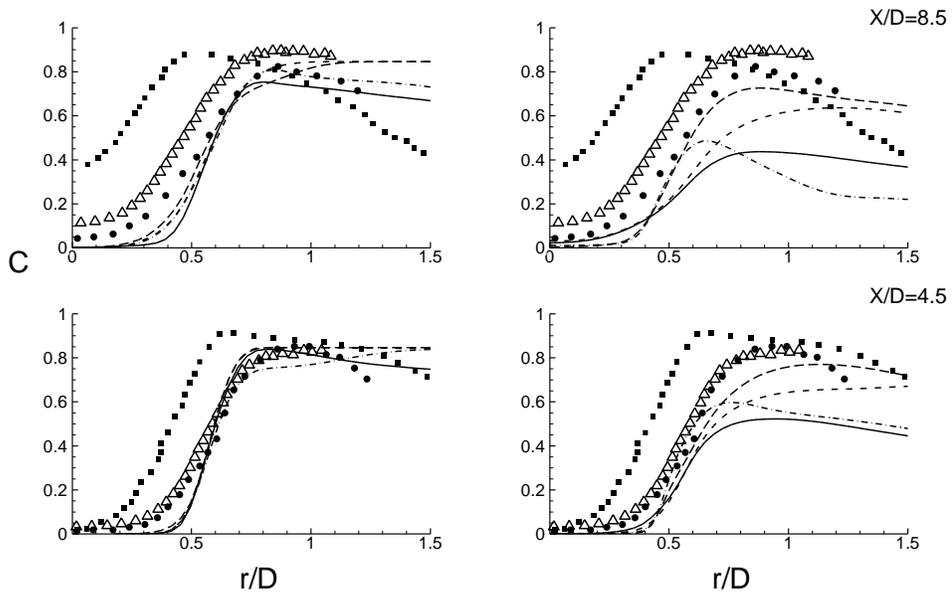

Figure 5. 2 step chemistry (left), 1 step chemistry (right): mean temperature C. Legend-See Fig. 4

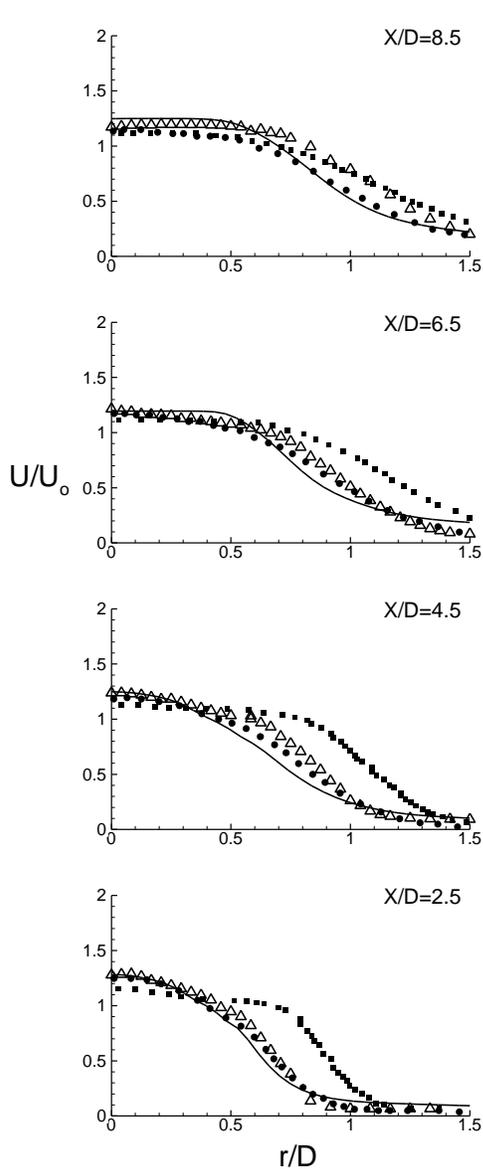

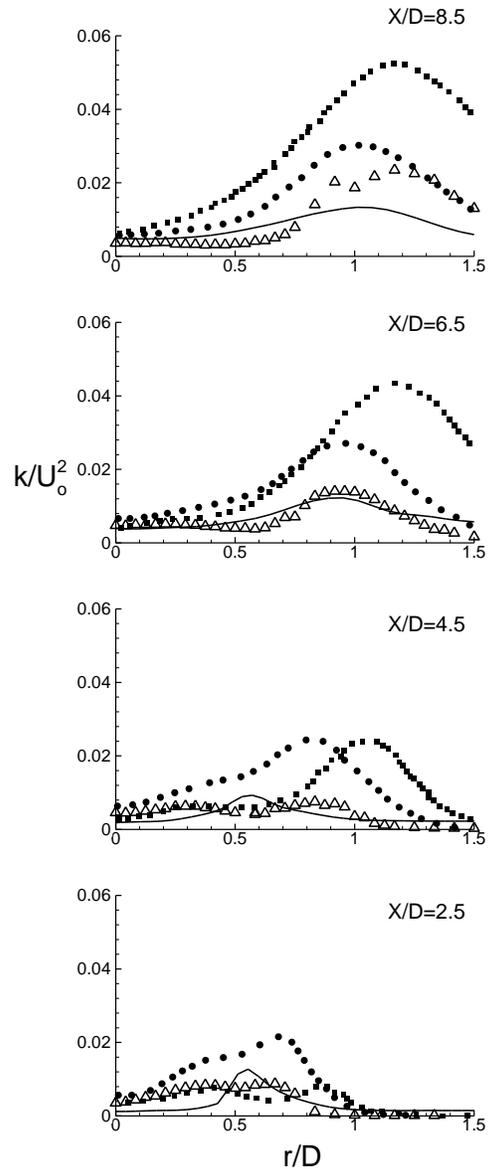

Figure 6. Reacting flow: Mean axial velocity $U/U_o$. Experimental data ($\Delta$), Lindstedt simulations ($\bullet$), Duchamp simulations ($\blacksquare$), TF model ( — ).

Figure 7. Reacting flow: turbulent kinetic energy $k/U^2_o$. Legend- See Fig. 6

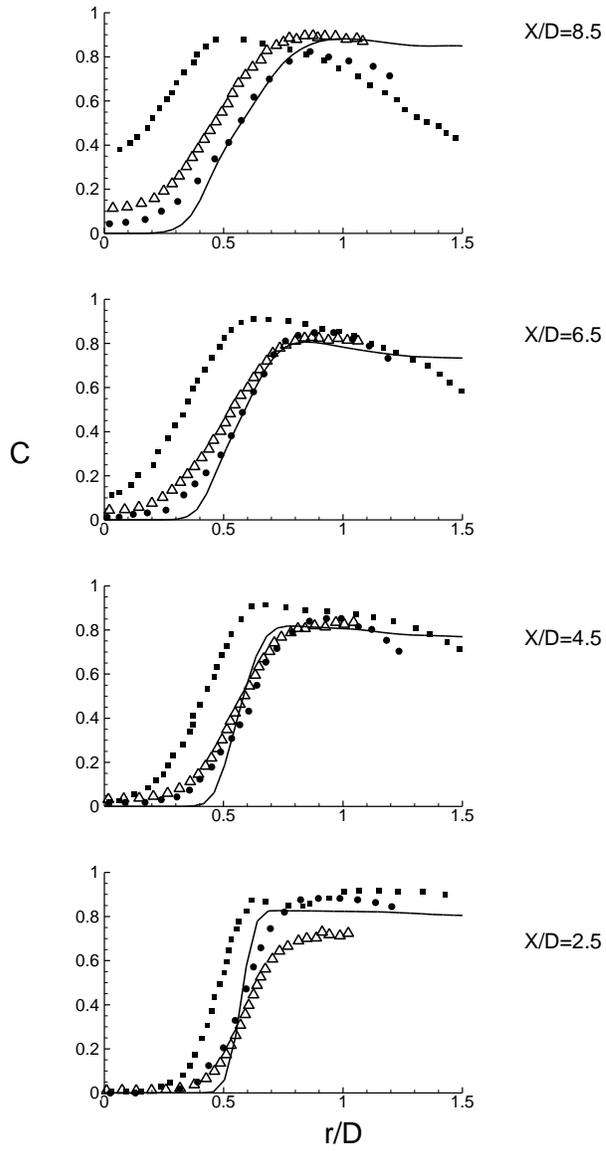

Figure 8. Reacting flow: Mean temperature C. Legend- See Fig. 6

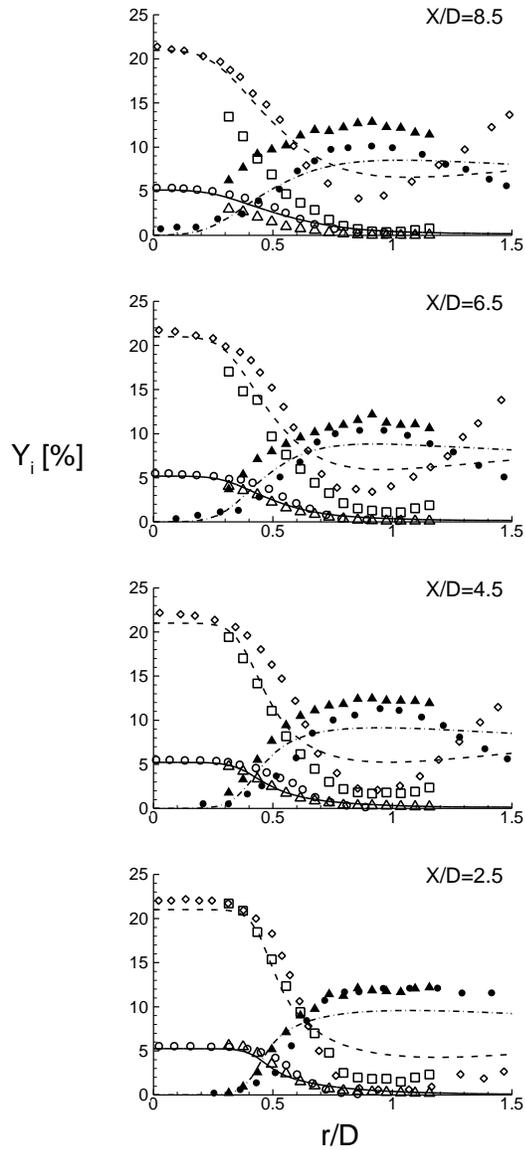

Figure 9. Reacting flow: Mean CH$_4$ (Δ, ○, □ ), O$_2$ (□, ◊, □ □ ), H$_2$O (▲, ●, □ . □ ) concentration. Experimental data (Δ, □, ▲), Lindstedt simulations (○, ◊, ●), Lines are TF model predictions.

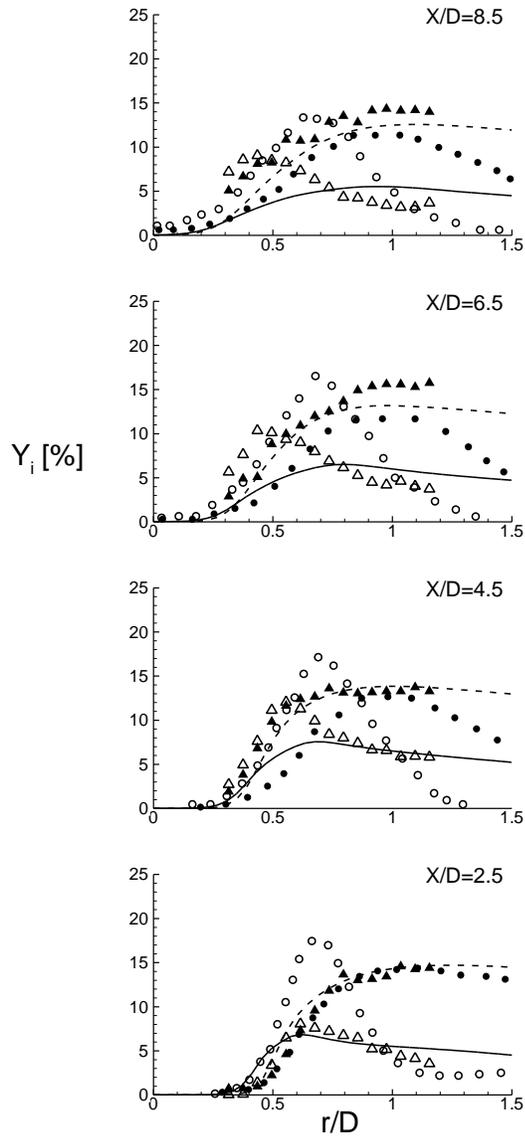

Figure 10. Reacting flow: Mean COx10 (Δ, ○, □ ), $CO_2$ (▲, ●, □ □ ) concentration. Experimental data (Δ, ▲), Lindstedt simulations (○, ●), Lines are TF model predictions.

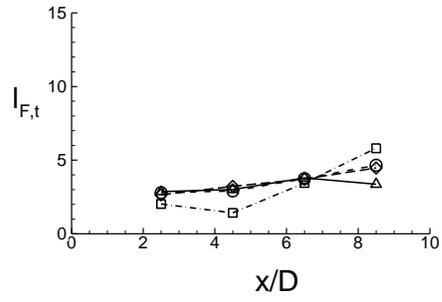

Figure 11. Turbulent flame-brush thickness $l_{f,t}$: Experimental data ($\Delta$), Lindstedt simulations (o), Duchamp simulations ($\square$), TF model ($\Diamond$).